\documentstyle[aps,12pt]{revtex}

\newcommand{\sect}[1]{\setcounter
        {equation}{0}\section{#1}}

\begin{document}
\draft

\title{Pseudo-orthogonal groups and
integrable dynamical systems \\ 
in two dimensions}

\author{Juan A.~Calzada}

\address{Dept. de Matem tica Aplicada a
la  Ingenier¡a, Universidad de Valladolid,
E-47011 Valladolid (Spain) 
\\ e-mail: juacal@wmatem.eis.uva.es}

\author{Mariano A. del Olmo}

\address{Dept. de F¡sica Te¢rica,
Universidad de Valladolid, 
E-47011 Valladolid (Spain)\\ 
e-mail: olmo@fta.uva.es}

\author{Miguel A.~Rodr¡guez}

\address{Dept. de F¡sica
Te¢rica, Universidad Complutense, E-28040
Madrid (Spain)\\ e-mail:
rodrigue@eucmos.sim.ucm.es}

%\date{\today}
\maketitle

\begin{abstract}
Integrable systems in low dimensions,  
constructed through the symmetry reduction
method, are studied using phase portrait and
variable separation techniques. In particular,
invariant quantities and explicit periodic
solutions are determined. Widely applied models
in Physics  are shown to appear as particular
cases of the method.
\end{abstract}

\noindent Running title: {\it Pseudo-orthogonal
groups and integrable systems}

\sect{Introduction}

Integrable Hamiltonian systems play a
fundamental role in the study and
description of physical systems, due to
their many interesting properties, both
from  the mathematical and physical points
of view. The construction of such models
represents a contribution to this field, and
many of them have proved to be of an
extraordinary physical interest. Let us
remind here the Morse \cite{Mo} and
P\"oschl-Teller \cite{PT} potentials in one
dimension, or the Calogero \cite{Ca,KK,Ms} and
Sutherland \cite{Su} potentials. These
constructions have been also considered from
many points of view. See for instance the
reviews \cite{Hi} and \cite{Pe}.

A method to construct these systems is the
use of the Marsden-Weinstein reduction
procedure \cite{MW}, or its extensions
\cite{GL}, to free Hamiltonians lying on a
$N$-dimensional homogeneous space under a
suitable Lie group. In this way (using an
appropriate momentum map), one assures the
integrability, or even the superintegrability 
\cite{Ev} of the system. In the first case,
there exist $N$ constants of motion in
involution, one of them the Hamiltonian. The
superintegrability requires more than $N$
constants of motion (not all of them in
involution) and more than one subset of $N$
constants in involution. There are good reasons
to suspect that any integrable system may be
constructed in this way, as a reduction of a
free one \cite{GL}, so the problem to construct
these systems and study their properties is a
profitable and very interesting field. A related
topic is the problem of separation of variables
for the associated Hamilton-Jacobi (HJ)
equations. As it is well known, the
existence of quadratic invariants allows to
classify and construct these systems
\cite{Ei,MP}, relating them in many
occasions to subgroups of the invariance
group.

A series of articles appeared in the last
years, has been devoted to the study of
these superintegrable systems constructed
using homogeneous spaces over the
pseudounitary groups $SU(p,q)$
\cite{OR1,OR2}. In particular, using the
maximal Abelian subalgebras (MASA) of the
corresponding algebras, one can build a
family of integrable systems of arbitrary
dimension, and  present their invariants and
the coordinate systems in which the HJ
equation is separated. The explicit
solutions and a unifying view of the compact
Cartan  subalgebra case have been presented
in \cite{CO1}.

Our aim in this article is to work in detail
the low dimensional cases. The reason is
twofold. On one side, the one-dimensional
case allows an easy geometric description of
the systems, through their phase portrait.
The potentials we obtain are not new, but
have been applied successfully in many
physical models (for instance, the
P\"oschl-Teller and Morse potentials). They
also appear in the study of quasi-exactly
solvable (QES) models \cite{Tu,GK}, as the case
of exactly solvable systems, providing examples
in which, from the quantum point of view, the
corresponding Schr\"odinger equation can be
solved algebraically (a part of the spectrum
for QES systems or an arbitrary number of
states for the exactly solvable ones). On
the other side, the 2-dimensional case can
be studied from the point of view of
variable separation, and we can solve the HJ
equation in a wide class of coordinate
systems, specially in the noncompact case
\cite{BK1,BK2}. The results we present here 
(in the 2- dimensional case) should be
considered in a local context. Considerations
about global behavior, which will differ from
the compact to the noncompact case, will not be
addressed in this work.

The article is organized as follows. In
Section 2 we present a concise description
of the method used to construct these
Hamiltonian systems. Section 3 is devoted to
the one-dimensional case, while the
2-dimensional case is studied in Section 4.
In each case we present the list of all the
conserved quantities for these systems in terms
of the generators of the corresponding algebras,
and the explicit form in the chosen coordinate
system. Conclusions and further outlook of this
research are discussed in Section 5.

\sect{Integrable Hamiltonian systems and
pseudounitary groups}

The results presented in this section are a
summary of the contents of \cite{OR1,OR2}.
We will include some of them to set the
notations which will be used in the
following sections.

We will consider the free Hamiltonian
($\mu,\nu=0,\ldots, N=p+q-1$):
\begin{equation} 
H=4cg^{\mu\bar{\nu}}p_{\mu}\bar{p}_{\nu}
\end{equation} 
(the bar denoting complex conjugate) defined in
the Hermitian hyperbolic space (with coordinates
$y^{\mu}\in {\bf C}$, satisfying
$g_{\bar{\mu}\nu}\bar{y}^{\mu}y^{\nu}=1$, and
conjugate momenta $p_{\mu}$):
\begin{equation}
SU(p,q)/SU(p-1,q)\times U(1)
\end{equation}
whose geometry is described in \cite{KN}. The
real constant $c$ is related to the sectional
curvature of the Hermitian space. See also
\cite{CO1} for a detailed analysis of this
space and its properties.

Using a maximal abelian subalgebra of
$su(p,q)$ \cite{OR3}, we carry a reduction
procedure \cite{MW}, in order to obtain a
reduced Hamiltonian (which is not free) in
the reduced space, a homogeneous $SO(p,q)$
space:
\begin{equation}
H=c\left({1\over2}g^{\mu\nu}p_{s^{\mu}}
p_{s^{\nu}}+V(s) \right)
\end{equation}
where $V(s)$ is a potential depending on the
real coordinates $s^{\mu}$. The set of complex
coordinates $y^{\mu}$ is transformed in the
reduction procedure into a set of ignorable
variables $x^{\mu}$ (which are the parameters
of the transformation associated to the MASA of
$u(p,q)$ used in the reduction) and the
coordinates $s^{\mu}$ with the constraint
$g_{\mu\nu}s^{\mu}s^{\nu}=1$.

If $Y_{\mu},\;\mu=0,\ldots,N$, is a basis of
the considered MASA of $u(p,q)$, formed by
pure imaginary matrices, the relation
between old ($y^{\mu}$) and new coordinates
($x^{\mu},s^{\mu}$) is:
\begin{equation} 
y^{\mu}=B(x)^{\mu}_{\nu}s^{\nu},\qquad
B(x)=\exp(x^{\mu}Y_{\mu})\label{matB}
\end{equation}
which assures the ignorability of the $x$
coordinates (the vector fields corresponding to
the MASA are straightened out in these
coordinates). The Jacobian matrix, $J$, is
easily obtained. If:
\begin{equation}
A^{\mu}_{\nu}={\partial y^{\mu}\over \partial
x^{\nu}}= (Y_{\nu})^{\mu}_{\rho} y^{\rho}
\end{equation}
then:
\begin{equation}
J={\partial (y,\bar{y})\over \partial (x,s)}=
\left(\begin{array}{cc} 
A & B\\ \bar{A} & \bar{B}
\end{array}\right).
\end{equation}

The Hamiltonian calculated in the new
coordinates is written as:
\begin{equation}
H=c\left({1\over2} g^{\mu\nu} p_{\mu}p_{\nu}+
V(s)\right),\qquad V(s)=
p_x^T(A^{\dagger}KA)^{-1} p_x\label{ham}
\end{equation}
where $p_x$ are the constant momenta associated
to the ignorable coordinates $x$ and $K$ is the
matrix defined by the metric $g$.

Note that, to obtain these Hamiltonians, we
need MASAs of $su(p,q)$ of dimension $N=p+q-1$
(corresponding to MASAs of $u(p,q)$ of dimension
$p+q$). We also require that these MASAs have a
representation in terms of imaginary
matrices that allows to write the
Hamiltonian in the form (\ref{ham}). Once we
have chosen a particular MASA, we can obtain
a set of invariants and also the
corresponding coordinate systems in which
the HJ  equation separates. The MASAs of
$su(p,q)$ are classified in \cite{OR3}, and
for low ranks are completely determined. The
corresponding potentials have been obtained
for  $SU(N)$ in \cite{BK1}, for $SU(2,1)$ in
\cite{BK2}, for $SU(2,2)$ in \cite{OR2} and
for any $SU(p,q)$, choosing as MASA one of the
Cartan subalgebras, in \cite{OR1}. From now
on, we will always use contravariant
coordinates, but we will write the indices as
subscripts in order to simplify the notation
and avoid the use of unnecessary brackets.

\sect{One-dimensional Hamiltonians}

One-dimensional Hamiltonians are always
integrable and the phase portrait gives a
complete description of the allowed motions.
We shall expose the main ideas in order to
achieve a better understanding of the more
complicate systems we will study in the next
section. We have two cases:
$su(2)$ and $su(1,1)$.

\subsection{$su(2)$}

We will use as a basis for $su(2)$ the operators
$X_1,X_2,X_3$, which are given in the natural
$2\times 2$ matrix representation by:
$$ X_1\rightarrow 
\left(\begin{array}{rr} 
i& 0 \\ 0 & -i
\end{array}\right),\qquad 
X_2\rightarrow
\left(\begin{array}{rr} 
0 &1 \\ -1 & 0
\end{array}\right),\qquad 
X_3\rightarrow
\left(\begin{array}{rr} 
0 &i\\ i& 0
\end{array}\right) $$
in the metric  $K={\mathop{\rm diag}}(1,1)$.

In the compact algebra $su(2)$ there is only
one class of MASAs, corresponding to the
Cartan subalgebra (CC) \cite{OR3}. A basis of a
representative of this class of MASA is:
\begin{equation} 
\left(\begin{array}{rr}
i& 0  \\ 0 & -i
\end{array}\right)
\end{equation}
A basis of the corresponding MASA of $u(2)$ can
be chosen as: 
\begin{equation} 
Y_0=
\left(\begin{array}{rr} 
i& 0 \\ 0 & 0 
\end{array}\right),
\qquad Y_1=
\left(\begin{array}{rr}
0 & 0 \\ 0 & i
\end{array}\right).
\label{matu2}
\end{equation}
 
The relation between old and new coordinates is
given by the matrix $B(x)$ in (\ref{matB}).
Note that a change of basis in the
corresponding MASA of $u(2)$ changes only the
parameters appearing in the potential:
\begin{equation}
y_0 = s_0 e^{ix_0},\qquad
y_1 = s_1 e^{ix_1}.\label{old}
\end{equation}

The Hamiltonian, following the general
expression (\ref{ham}), is:
\begin{equation}
H=c\left[{1\over 2} (p_{s_0}^2+ p_{s_1}^2)+
V(s)\right],\qquad V(s)={m_0^2\over
s_0^2}+{m_1^2\over s_1^2}\label{cartham}
\end{equation}
with the constraint $s_0^2+s_1^2=1$.
Parameterizing the circle $S^1$ in spherical
coordinates: $s_0=\cos \phi,\; s_1=\sin\phi$, 
we get the Hamiltonian ($c=1$):
\begin{equation}
H(\phi)={1\over 2} p_{\phi}^2+V(\phi),\qquad
V(\phi)={m_0^2\over \cos^2\phi}+{m_1^2\over
\sin^2\phi}.\label{pot3}
\end{equation}

We have only one second order conserved quantity,
the Hamiltonian, which is equal to the Casimir
of the algebra, $C$, up to an additive constant:
\begin{equation}
\hat{Q}_1=X_2^2+X_3^2.
\end{equation}
The square of the generator in the compact Cartan
subalgebra, $C_1=X_1^2$, is constant
after the reduction and $C=C_1+\hat{Q}_1$.

The specific values of the real positive
constants $m_0,m_1$ play no essential role in
the qualitative description of the orbits and
trajectories of this system. The potential has 
singularities (in the generic case) in
$\phi=0,\pi/2,\pi,3\pi/2$. When $m_0$ or $m_1$
are equal to zero  we have only two
singularities in $0,\pi$ or $\pi/2,3\pi/2$,
respectively.

The particles are confined inside a sector,
and there, the motion is periodic, with an
equilibrium point (a center in the phase
space) corresponding to the unique minimum
of the potential, in $\tan\phi=\sqrt{m_1/m_0}$.
The solution can be easily computed, using 
Hamilton equations. The potential is bounded
from below, and the energy is always positive
($E\geq (m_0+m_1)^2$). Though the use of HJ
equation is not necessary in this context of
one-dimensional systems, we will write down
the equation in order to compare with the
two dimensional cases. In fact, when we will
make separation of variables there, we will
find again this equation:
\begin{equation}
{1\over 2}\left({\partial
S\over \partial \phi}\right)^2 +{m_0^2\over
\cos^2\phi} +{m_1^2\over
\sin^2\phi}= E\label{hjsu2}
\end{equation}
with solution ($u=\cos^2\phi$):
\begin{equation}
u={1\over 2E}\left(b+\sqrt{b^2-4m_0^2E}\cos
2\sqrt{2E}t\right)
\end{equation}
and $b=m_0^2-m_1^2+E$. This solution is
obviously much simpler to find if we consider
the equation of orbits in the phase portrait of
this system.

For instance, if  $m_0=0,\;m_1=1$, the
solution is:
\begin{equation} 
s_0=\cos\phi =\sqrt{1-{1\over E}}\cos
\sqrt{2E}t
\end{equation}
that is, a system with similar solutions to a
harmonic oscillator, but now the frequency
depends on the energy.

\subsection{ $su(1,1)$}

The noncompact algebra $su(1,1)$ has three
nonconjugate classes of MASAs, compact
Cartan subalgebra, noncompact Cartan
subalgebra and a class of nilpotent maximal
abelian subalgebras, (MANS) \cite{OR3}. We
will fix the metric to be:
\begin{equation}
K=\left(\begin{array}{rr} 1 & 0 \\ 0 
& -1 \end{array}\right).\label{metu11}
\end{equation}
and the basis $\{X_1,X_2,X_3\}$ is given in
the $2\times 2$ matrix
representation through the correspondence:
$$ X_1\rightarrow 
\left(\begin{array}{rr} 
i& 0 \\ 0 & -i
\end{array}\right),\qquad 
X_2\rightarrow
\left(\begin{array}{rr} 
0 &1 \\ 1 & 0
\end{array}\right),\qquad 
X_3\rightarrow
\left(\begin{array}{rr} 
0 &i\\ -i& 0
\end{array}\right) $$
\medskip

{\it I. Compact Cartan subalgebra (CC)}
\smallskip

We choose, as representative of this class,
the same matrices as in (\ref{matu2}).
Hence, the old and new coordinates are
related in the same way they did in the
$su(2)$ case (\ref{old}), and the
Hamiltonian is now:
\begin{equation}
H=c\left[{1\over2} (p_{s_0}^2- p_{s_1}^2)+
V(s)\right],\qquad V(s)={m_0^2\over
s_0^2}-{m_1^2\over s_1^2}
\end{equation}
with the constraint $s_0^2-s_1^2=1$. This
hyperbola can be described with a coordinate
$\phi$ varying in the real line:
$s_0=\cosh \phi,\; s_1=\sinh \phi$, and the
Hamiltonian in these coordinates is ($c=-1$):
\begin{equation}
H(\phi)={1\over 2}
p_{\phi}^2+V(\phi),\qquad V(\phi)=-{m_0^2\over
\cosh^2 \phi}+{m_1^2\over
\sinh^2 \phi}.\label{pot1}
\end{equation}

The second order invariant (the Hamiltonian) is:
\begin{equation}
\hat{Q}_1=X_2^2+X_3^2
\end{equation}
and the trivial constant associated to the
MASA is $C_1=X_1^2$. Hence, the Casimir in terms
of these two quantities is $C=C_1-\hat{Q}_1$.

The HJ equation is:
\begin{equation}
{1\over 2}\left({\partial S\over \partial
\phi}\right)^2 -{m_0^2\over \cosh^2\phi}
+{m_1^2\over \sinh^2\phi}=E
\end{equation}
and the solution depends on the values of
$E$ and the parameters.

Considering different values of the
parameters $m_0,m_1$ we obtain three different
systems. 

a) If $m_1 \neq 0$ the potential has a
singularity in $\phi=0$. It is easy to check
that, if $m_1\geq m_0$, there are no minima for
the potential and all the motions are unbounded
(with a turning point). The energy is always
positive. The parameters $m_0,m_1$ do not
modify qualitatively the phase portrait or
the form of the solutions. The solution can
be written as ($u=\cosh^2\phi$):
\begin{equation}
u={1\over 2E} \left(-b+ \sqrt{b^2+4m_0^2E} 
\cosh 2\sqrt{2E}t\right) \label{u11}
\end{equation}
and $b=m_0^2-m_1^2-E$. If $m_0=0,m_1=1$, the
solution is:
\begin{equation}
s_0(t)=\sqrt{1+{1\over E}}\cosh \sqrt{2E}t.
\end{equation}

b) If $m_0>m_1>0$, the potential has two minima,
symmetric respect to the origin where it has
the singularity. The energy is bounded from
below: $E\geq -(m_0-m_1)^2$, the value
$E=-(m_0-m_1)^2$ corresponding to the
equilibrium solution in the center of the
phase space. The other solutions are easily
calculated ($b$ has the same value as in
case a)):

\noindent i) $-(m_0-m_1)^2<E<0$
\begin{equation}
s_0={1\over \sqrt{2|E|}}\left[b+
\sqrt{b^2+4Em_0^2}\cos
2\sqrt{2|E|}t\right]^{1/2}.
\end{equation}  
ii) $E=0$ 
\begin{equation}
s_0=\left[{m_0^2\over
m_0^2-m_1^2}+2(m_0^2-m_1^2) t^2\right]^{1/2}.
\end{equation}
When $E>0$ we get the solution (\ref{u11}).

c) If $m_1=0$ there is no singularity in the
potential, which has a minimum in $\phi=0$,
with periodic motions of negative energy and
unbounded motions of positive or zero energy.
The multiplicative constant $m_0$ plays no
essential role for the qualitative description
of the system. The solutions can be read off
from case b) with $m_1=0$.

\medskip

{\it II. Noncompact Cartan subalgebra (NC)}
\smallskip

A representative subalgebra of this class
has the basis (in the  metric (\ref{metu11})):
\begin{equation} 
Y_1=\left(\begin{array}{rr} 0 
& i\\ -i& 0
\end{array}\right)\label{basesu11}
\end{equation}
and we will add the matrix $Y_0=i I$ to get a
MASA  of $u(1,1)$. The new and old coordinates
are related in a slightly more complicated way:
\begin{equation}
y_0=e^{ix_0}(s_0\cosh x_1+is_1\sinh x_1),\qquad
y_1=e^{ix_0} (-is_0\sinh x_1+s_1\cosh x_1)
\end{equation}
and the Hamiltonian is written in the new
coordinates as:
\begin{equation}
 H=c\left[{1\over 2} (p_{s_0}^2- p_{s_1}^2) +
V(s)\right],\qquad V(s)={m_0^2-m_1^2+4m_0m_1
s_0s_1\over 1+4s_0^2s_1^2}
\end{equation} 
and the constraint $s_0^2-s_1^2=1$. Using again
the $\phi$ coordinate as in the previous case,
we get:
\begin{equation}
H(u)={1\over 2} p_{\phi}^2+V(\phi),\qquad
V(\phi)=-{m_0^2-m_1^2 + 2m_0m_1 \sinh 2\phi
\over \cosh^2 2\phi}.
\end{equation}

The Casimir is written as $C=\hat{Q}_1-C_1$
where 
$C_1=X_3^2$, the square of the generator of the
noncompact Cartan subalgebra, and
\begin{equation}
\hat{Q}_1=X_1^2-X_2^2
\end{equation}
which is equal to the Hamiltonian.

We will also write down HJ equation for future
references: 
\begin{equation} {1\over
2}\left({\partial S\over \partial \phi}\right)^2
-{m_0^2-m_1^2 + 2m_0m_1 \sinh 2\phi \over
\cosh^2 2\phi}=E.\label{hjnc} 
\end{equation}

If $m_0$ or $m_1$ are equal to 0, we obtain
similar results to those of the case of
compact Cartan MASA, an attractive potential
if $m_0^2>m_1^2$ and a repulsive one in the
opposite case. If $m_0m_1\neq 0$, the
potential is qualitatively the same for all
values of $m_0$ and $m_1$.

The potential  $V(\phi)$ has two extrema in
$\sinh 2\phi=m_1/m_0,-m_0/m_1$. The first
point corresponds to a minimum (a center in the
phase portrait), and
the potential takes the value $V=-m_0^2$. The
second point is a maximum (a saddle point in the
phase portrait), and $V=m_1^2$ there. The energy
is bounded from below ($E\geq -m_0^2$) and the
explicit solutions are ($u=\sinh 2\phi$):

\noindent i) $-m_0^2<E<0$
\begin{equation}
u={1\over |E|}\left[m_0m_1+
\sqrt{(E+m_0^2)(m_1^2-E)}
\cos 2\sqrt{2|E|}t\right].\label{hjnc1}
\end{equation}
ii) $E=0$
\begin{equation}
u=-{m_0^2-m_1^2\over
2m_0m_1}+4m_0m_1t^2.
\end{equation}  
iii) $0<E<m_1^2$
\begin{equation}
u={1\over E} \left[-m_0m_1+
\sqrt{(E+m_0^2)(m_1^2-E)}
\cosh 2\sqrt{2|E|}t\right].
\end{equation} 
iv) $E=m_1^2$
\begin{equation}
u=-m_0+e^{2\sqrt{2}m_1t}.
\end{equation} 
v) $E>m_1^2$
\begin{equation}
u={1\over E}\left[-m_0m_1+
\sqrt{(E+m_0^2)(E-m_1^2)}\sinh
2\sqrt{2E}t\right].\label{hjnc5}
\end{equation}

\medskip

{\it III. Nilpotent subalgebra (NIL)}
\smallskip

Though the simplest representative of this
class of subalgebras is obtained in the
skewdiagonal metric, we will use again the
diagonal one, because in this way, the
kinetic term is also diagonal. We will take
as a basis:
\begin{equation}
Y_1=
\left(\begin{array}{rr} 
i& i\\ -i& -i
\end{array}\right)
\end{equation}
which is a nilpotent matrix. As in the
noncompact case we will also use $Y_0=iI$ to
complete the basis of a $u(1,1)$ MASA. Old and
new coordinates satisfy: 
\begin{equation}
y_0=e^{ix_0} ( ( 1 + ix_1 ) s_0 + ix_1
s_1 ),\qquad  
y_1=e^{ix_0} ( -ix_1 s_0 + (1 - ix_1 )
s_1).
\label{nilsu11}
\end{equation}

The Hamiltonian is:
\begin{equation}
H=c\left[{1\over 2} (p_{s_0}^2-
p_{s_1}^2)+V(s)\right], \qquad 
V(s)={2m_0m_1\over (s_0+s_1)^2}-{m_1^2\over
(s_0+s_1)^4}
\end{equation}
with the constraint (which is the same for
all the subalgebras in the $su(1,1)$ case,
as we are using the same metric):
$s_0^2-s_1^2=1$. The expression of the
Hamiltonian in terms of the $\phi$ coordinate
($c=-1$) is: 
\begin{equation} H(\phi)={1\over 2}
p_{\phi}^2+V(\phi),\qquad V(\phi)=m_1^2
e^{-4\phi} - 2m_0m_1 e^{-2\phi}.\label{pot2}
\end{equation} 

The Hamiltonian in terms of the second order
operators in the enveloping algebra is again
($\{X_i,X_j\}=X_iX_j+X_jX_i$):
\begin{equation}
\hat{Q}_1=2X_1^2+\{X_1,X_3\}-X_2^2
\end{equation}
and the trivial constant $C_1$ is
equal to $(X_1+X_3)^2$, with $C=\hat{Q}_1-C_1$.

The HJ equation is:
\begin{equation}{1\over 2}\left({\partial
S\over \partial
\phi}\right)^2  - 2m_0m_1 e^{-2\phi} + m_1^2
e^{-4\phi}=E.\label{nileq}
\end{equation}

We will assume $m_1\neq 0$ to obtain nontrivial
results. Depending on the constants $m_0,m_1$ 
we get essentially two classes of systems:

1) If $m_0m_1\leq 0$ there is no extremum
for the potential, and the energy is always
positive. The solutions, with a unique
turning point, are given by:
\begin{equation}
e^{2\phi}={m_1\over E}\left[-m_0+
\sqrt{E+m_0^2}\cosh
\sqrt{2E} t\right].\label{nil}
\end{equation}

2) If $m_0m_1>0$ the
potential has a minimum, in 
$\phi=(1/2)\log (m_1/m_0)$ and the energy
is bounded from below, $E>-m_0^2$. As in the
first case, the values of the parameters are
not essential if they satisfy the constraints.
The solutions for the different values of the
energy are:

\noindent i) $-m_0^2<E<0$
\begin{equation}
e^{2\phi}={m_1\over
|E|}\left[m_0+\sqrt{E+m_0^2}\cos
\sqrt{2E} t\right].
\end{equation} 
ii) $E=0$
\begin{equation}
e^{2\phi}={m_1\over 2m_0}+ 4m_0 m_1 t^2.
\end{equation}
If $E>0$ the solution is the same as
(\ref{nil}).

This case completes the set of one-dimensional
Hamiltonians obtained through a reduction
procedure out of free Hamiltonians,
invariant under $SU(p,q),\; p+q=2$, and
defined over a homogeneous space of the
corresponding group. In Table I, we
present a summary of these Hamiltonians in
the one dimensional case.

In the next section we will treat the
2-dimensional Hamiltonians associated to the
rank 2 algebras $su(3)$ and
$su(2,1)$.

\sect{The 2-dimensional case}

There are two pseudounitary algebras to be
used to construct superintegrable
Hamiltonians of dimension 2,
$su(3)$ and $su(2,1)$. We will treat
separately both cases.

\subsection{$su(3)$}

We will use as a basis for $su(3)$ the operators
$\{X_1,\ldots,X_8\}$ which are given in
the $3\times 3$ matrix representation by:
\begin{eqnarray*} X_1\rightarrow 
\left(\begin{array}{rrr}  
i& 0 & 0 \\ 
0 &-i& 0 \\ 
0 & 0 & 0
\end{array}\right),& 
X_2\rightarrow 
\left(\begin{array}{rrr}  
0 & 0 & 0 \\ 
0 & i& 0 \\ 
0 & 0 &-i
\end{array}\right),& 
X_3\rightarrow 
\left(\begin{array}{rrr}  
0 & 1 & 0 \\ 
-1 & 0 & 0 \\ 
0 & 0 & 0
\end{array}\right),\\
X_4\rightarrow 
\left(\begin{array}{rrr}  
0 & i& 0 \\ 
i& 0 & 0 \\ 
0 & 0 & 0
\end{array}\right),& 
X_5\rightarrow
\left(\begin{array}{rrr}  
0 & 0 & 1 \\ 
0 & 0 & 0 \\ 
-1 & 0 & 0
\end{array}\right),& 
X_6\rightarrow 
\left(\begin{array}{rrr}
0 & 0 & i\\ 
0 & 0 & 0 \\
i& 0 & 0
\end{array}\right),\\
X_7\rightarrow 
\left(\begin{array}{rrr}
0 & 0 & 0 \\ 
0 & 0 & 1 \\ 
0 &-1 & 0
\end{array}\right),& 
X_8\rightarrow
\left(\begin{array}{rrr}
0 & 0 & 0\\ 
0 & 0 & i\\ 
0 & i& 0
\end{array}\right).\\
\end{eqnarray*}
when the metric is: $K={\mathop{\rm
diag}}(1,1,1)$.

In the compact case there is only one MASA,
the Cartan subalgebra, generated by the
matrices:
\begin{equation}
\left(\begin{array}{rrr} 
i\\ & -i\\ & & 0
\end{array}\right),\qquad
\left(\begin{array}{rrr} 
0 \\ & i\\ & & -i
\end{array}\right)
\end{equation} 
and we shall use the following basis for the
corresponding MASA in $u(3)$:
\begin{equation}
Y_0=\left(\begin{array}{rrr} 
i\\ & 0 \\ & & 0
\end{array}\right),\qquad
Y_1=\left(\begin{array}{rrr}
0 \\ & i\\ & & 0
\end{array}\right),\qquad
Y_2=\left(\begin{array}{rrr} 
0 \\ & 0 \\ & &
i\end{array}\right).\label{basesu3}
\end{equation}

The coordinates $s$ are related to the
coordinates $y$ in the same way as in
$su(2)$, (\ref{old}):
\begin{equation}
y_0 = s_0 e^{ix_0},\qquad 
y_1 = s_1 e^{ix_1},\qquad 
y_2 = s_2 e^{ix_2}\label{coordsu3}
\end{equation}
and the Hamiltonian has also the same form of
all cases using compact Cartan subalgebras
(\ref{cartham}):
\begin{equation}
H={1\over 2}\left(
p_{0}^2+p_1^2+p_2^2\right)+V(s),\qquad
V(s)={m_0^2\over s_0^2}+{m_1^2\over s_1^2}+
{m_2^2\over s_2^2}
\end{equation}
with the constraint $s_0^2+s_1^2+s_2^2=1$.

In the one-dimensional case, there was only
one invariant, which was the Hamiltonian.
In this case, we can construct three
invariants, only two of them in involution at
the same time  (one of them the Hamiltonian).
The system is superintegrable in the sense of
\cite{Ei}. These invariants are \cite{OR1}:
\begin{eqnarray}
R_{01}&=&(s_0p_1-s_1p_0)^2+
\left( m_0{s_1\over s_0}+m_1{s_0\over
s_1}\right)^2,\nonumber\\
R_{02}&=&(s_0p_2-s_2p_0)^2+
\left( m_0{s_2\over s_0}+m_2{s_0\over
s_2}\right)^2,\label{invsu3}\\
R_{12}&=&(s_1p_2-s_2p_1)^2+
\left( m_1{s_2\over s_1}+m_2{s_1\over
s_2}\right)^2.\nonumber
\end{eqnarray} 

The sum of these three invariants is the
Hamiltonian up to an additive constant. In
order to study the solutions of this problem
we need construct a coordinate system in
which the corresponding HJ equation
separates into a system of ordinary
differential equations. As in \cite{CO1} we will
use spherical coordinates \cite{BK1}, defined
by:
\begin{equation}
s_0 = \cos\phi_2\cos \phi_1,\qquad 
s_1 = \cos\phi_2\sin \phi_1,\qquad 
s_2 = \sin \phi_2.
\end{equation}
and the Hamiltonian is written as ($c=1$):
\begin{eqnarray}
H&=&{1\over 2}\left( 
p_{\phi_2}^2+{p_{\phi_1}^2\over
\cos^2 \phi_2}\right)+
V(\phi_1,\phi_2)\nonumber\\
V(\phi_1,\phi_2)&=&
{1\over \cos^2\phi_2}\left({m_0^2\over
\cos^2\phi_1}+{m_1^2\over
\sin^2\phi_1}\right) + {m_2^2\over
\sin^2\phi_2}
\end{eqnarray} 
where the constants $m_0,m_1,m_2$ are chosen to
be nonnegative. 

The second order conserved quantities
(\ref{invsu3}) (we will follow the
notation $\hat{Q}$ for these operators) can be
written in terms of the basis
$\{X_1,\ldots,X_8\}$:
\begin{equation}\hat{Q}_1=X_3^2+X_4^2,\qquad
\hat{Q}_2=X_5^2+X_6^2,\qquad
\hat{Q}_3=X_7^2+X_8^2 
\end{equation}
with commutation relations (the commutator is a
third order element which plays no essential
role in the method):
$$ [\hat{Q}_1,\hat{Q}_2]=[\hat{Q}_2,
\hat{Q}_3]=[\hat{Q}_3,\hat{Q}_1] $$
The Casimir is:
$$C=4C_1+2C_2+4C_3+3\hat{Q}_1+3
\hat{Q}_2+3\hat{Q}_3$$
where
$$ C_1=X_1^2,\qquad C_2=\{X_1,X_2\},\qquad
C_3=X_2^2 $$
are the second order operators in the
enveloping algebra of the compact Cartan
subalgebra.

The Hamiltonian is
$$ H= Q_1+Q_2+Q_3+{\rm constant} $$
where $Q_i$ is the expression of $\hat{Q}_i$ in
spherical coordinates \cite{BK1}:
\begin{eqnarray*}
Q_1&=&{1\over
2}p_{\phi_1}^2+{m_0^2\over
\cos^2\phi_1}+ {m_1^2\over
\sin^2\phi_1} \\
Q_2&=& \tan^2 \phi_2\left({1\over
2}p_{\phi_1}^2
\sin^2\phi_1 +{m_0^2\over
\cos^2 \phi_1}\right)+\cos^2 \phi_1
\left({1\over 2}p_{\phi_2}^2+{m_2^2\over
\tan^2 \phi_2}\right)\\
&&+ {1\over 2}p_{\phi_1}p_{\phi_2}\sin 2\phi_1
\tan \phi_2\\
Q_3&= &\tan^2 \phi_2\left({1\over
2}p_{\phi_1}^2
\cos^2\phi_1 +{m_1^2\over
\sin^2 \phi_1}\right)+\sin^2 \phi_1
\left({1\over 2}p_{\phi_2}^2+{m_2^2\over
\tan^2 \phi_2}\right)\\
&&- {1\over 2}p_{\phi_1}p_{\phi_2}\sin 2\phi_1
\tan\phi_2
\end{eqnarray*}

The HJ equation is: 
\begin{equation}
{1\over 2} \left( {\partial S\over \partial
\phi_2}\right)^2 +  {m_2^2\over
\sin^2\phi_2}+
{1\over \cos^2\phi_2}\left({1\over 2}  \left(
{\partial S\over
\partial \phi_1}\right)^2+{m_0^2\over
\cos^2\phi_1} + {m_1^2\over
\sin^2\phi_1}\right)=E
\end{equation}
and separates into two ordinary
differential equations using
$S(\phi_1,\phi_2)=S_1(\phi_1)+ S_2(\phi_2)-Et$:
\begin{eqnarray}
{1\over 2} \left( {\partial
S_1\over \partial
\phi_1}\right)^2+{m_0^2\over
\cos^2\phi_1} + {m_1^2\over
\sin^2\phi_1} & = & \alpha_1,\label{hjsu3}\\
{1\over 2} \left( {\partial S_2\over \partial
\phi_2}\right)^2 +  {m_2^2\over
\sin^2\phi_2}+  {\alpha_1\over \cos^2\phi_2}
& = & \alpha_2
\end{eqnarray}
where $\alpha_2=E$ and $\alpha_1$ are the
separation constants (which are positive).
These equations have the same form as those in
(\ref{hjsu2}) The solutions are easily computed
and can be found as particular cases in
\cite{CO1}. The potential has singularities
along the coordinate lines:
$\phi_1=0,\pi/2,\pi,3\pi/2$, and
$\phi_2=\pi/2,3\pi/2$ in the generic case.
It has a unique minimum inside each
regularity domain. An analysis of the
associated dynamical system (Hamilton
equations) shows that all the orbits in a
neighborhood of the critical point (center)
are closed and hence, the corresponding
trajectories are periodic (a direct consequence
of the correspondence between extrema of the
potential and critical points of the phase
space).  Let us restrict to the domain
$0<\phi_1,\phi_2<\pi/2$, where the minimum
is in $\tan\phi_1=\sqrt{m_1/m_0},\;\tan\phi_2=
\sqrt{m_2/(m_0+m_1)}$. The value of the
potential at this point is
$(m_0+m_1+m_2)^2$, hence the energy
$E$ is  bounded from below
($E\ge (m_0+m_1+m_2)^2$). The explicit
solutions are:
\begin{eqnarray}
\cos^2\phi_2 & = & {1\over 2E}\left[
b_2+\sqrt{b_2^2-4\alpha_1 E}\cos
2\sqrt{2E}t\right], \\
\cos^2\phi_1 & =  &{1\over 2\alpha_1}\left[
b_1+{1\over\cos^2\phi_2}
\left[{b_1^2-4\alpha_1 m_0^2\over 
b_2^2-4\alpha_1E}\right]^{1/2}\left(
(b_2\cos^2\phi_2-2\alpha_1)\sin
2\sqrt{2\alpha_1}\beta_1\right.\right.
\nonumber\\ 
&&\left.\left.
+2\sqrt{\alpha_1}[(b_2
-E\cos^2\phi_2)\cos^2\phi_2-\alpha_1]^{1/2}
\cos2\sqrt{2\alpha_1}\beta_1\right)\right]
\end{eqnarray}
where $b_1=\alpha_1+m_0^2-m_1^2$ and
$b_2=E+\alpha_1-m_2^2$.

Let us remark that these results reflect
essentially the case $su(2)$. In fact all
systems we can construct using Cartan
subalgebras can be described in a unified
way as it was shown in \cite{OR1,CO1}. 

\subsection{$su(2,1)$}

The basis we will use is formed by the set of
operators $\{X_1,\ldots,X_8\}$ which are given
in the $3\times 3$ matrix representation by:
\begin{eqnarray*} X_1\rightarrow 
\left(\begin{array}{rrr}  
i& 0 & 0 \\ 
0 &-i& 0 \\ 
0 & 0 & 0
\end{array}\right),& 
X_2\rightarrow 
\left(\begin{array}{rrr}  
0 & 0 & 0 \\ 
0 & i& 0 \\ 
0 & 0 &-i
\end{array}\right),& 
X_3\rightarrow 
\left(\begin{array}{rrr}  
0 & 1 & 0 \\ 
-1 & 0 & 0 \\ 
0 & 0 & 0
\end{array}\right),\\
X_4\rightarrow 
\left(\begin{array}{rrr}  
0 & i& 0 \\ 
i& 0 & 0 \\ 
0 & 0 & 0
\end{array}\right),& X_5\rightarrow
\left(\begin{array}{rrr}  
0 & 0 & 1 \\ 
0 & 0 & 0 \\ 
1 & 0 & 0
\end{array}\right),& X_6\rightarrow 
\left(\begin{array}{rrr}  
0 & 0 & i\\ 
0 & 0 & 0 \\
-i& 0 & 0
\end{array}\right),\\
X_7\rightarrow 
\left(\begin{array}{rrr} 
0 & 0 & 0 \\ 
0 & 0 & 1 \\ 
0 & 1 & 0
\end{array}\right),& X_8\rightarrow
\left(\begin{array}{rrr}  
0 & 0 & 0\\ 
0 & 0 & i\\ 
0 &-i& 0
\end{array}\right).
\end{eqnarray*}

According to general results \cite{OR3},
$su(2,1)$ has four MASAs, two of them Cartan
subalgebras (compact and noncompact), one
orthogonally decomposable subalgebra, with one
nilpotent element, and one nilpotent subalgebra.
We will discuss these four cases in the
following. Although some of these subalgebras
have a simpler expression in some skewdiagonal
metrics, we will always use the diagonal one:
\begin{equation} K=
\left(\begin{array}{rrr} 
1 \\ & 1\\ & & -1
\end{array}\right)
\end{equation} 
because, in this way, the kinetic part is always
diagonal. There are nine coordinate systems
associated to $O(2,1)$ free Hamiltonians:
spherical, hyperbolic, elliptic (I and II),
complex elliptic, horospheric, elliptic
parabolic, hyperbolic parabolic and semicircular
parabolic \cite {BK2,Ol}. Not all of them
will separate our systems because these are
not free. However, the appropriate systems
have been computed in \cite{BK2} and we will
use their results.

\vfill
\newpage

{\it I. Compact Cartan subalgebra (CC)}
\smallskip

The compact Cartan subalgebra has a basis
formed by the same two matrices we used in
$su(3)$, and the same situation happens for
the corresponding MASA in $u(2,1)$
(\ref{basesu3}). The coordinates $s$ are
also related to the coordinates $y$ as they
did in the compact case $su(3)$
(\ref{coordsu3}).

However, the Hamiltonian reflects the
noncompact character of $su(2,1)$:
\begin{equation}
H=c\left({1\over 2}\left(
p_{0}^2+p_1^2-p_2^2\right)+V(s)\right),\qquad
V(s)={m_0^2\over s_0^2}+{m_1^2\over s_1^2}-
{m_2^2\over s_2^2}
\end{equation}
where the constraint $s_0^2+s_1^2-s_2^2=1$ must
be satisfied.

This Hamiltonian separates in four
coordinate systems, spherical, hyperbolic
and elliptic I and II \cite{BK2}. We will
use spherical coordinates to discuss the
explicit solution.
\begin{equation} 
s_0=  \cosh \phi_2 \cos \phi_1,\qquad 
s_1=  \cosh \phi_2 \sin \phi_1,\qquad 
s_2=  \sinh \phi_2. 
\end{equation} 
Choosing $c=-1$, we have the
Hamiltonian in these coordinates:
\begin{eqnarray} 
H & = & {1\over 2}
\left( p_{\phi_2}^2-{p_{\phi_1}^2\over
\cosh^2 \phi_2}\right)
+V(\phi_1,\phi_2),\nonumber \\
V(\phi_1,\phi_2) & = &  -{1\over
\cosh^2\phi_2}\left({m_0^2\over
\cos^2\phi_1}+  {m_1^2\over
\sin^2\phi_1}\right)+ {m_2^2\over
\sinh^2\phi_2}.
\label{hamcom}\end{eqnarray}

Due to the form of the potential the
constants $m_0,m_1,m_2$ can be chosen
nonnegative. The potential is regular inside
the domain $0<\phi_1<\pi/2$, $0<\phi_2<\infty$.
It has a saddle point: $\tan
\phi_1=\sqrt{m_1/m_0}$, $\tanh\phi_2=\sqrt{m_2/
(m_0+m_1)}$, if $m_0+m_1>m_2$.
However, due to the special form of the kinetic
term (which is not  positive definite), it is
easy to check that the associated dynamical
system has all the orbits in a neighborhood of
the critical point (which is also a center as
in the compact case) closed and again, the
corresponding trajectories are periodic.  

The second order operators in the enveloping
algebra of this MASA are:
$$ C_1=X_1^2,\qquad C_2=\{X_1,X_2\},\qquad
C_3=X_2^2$$ 
The quadratic constants of motion lying in the
enveloping algebra of $su(2,1)$ and commuting
with the elements in the compact Cartan
subalgebra are:
\begin{equation} \hat{Q}_1=X_3^2+X_4^2,\qquad
\hat{Q}_2=X_5^2+X_6^2,\qquad
\hat{Q}_3=X_7^2+X_8^2 \end{equation}
with commutation relations:
$$ [\hat{Q}_1,\hat{Q}_2]=[\hat{Q}_3,\hat{Q}_2]=
-[\hat{Q}_3,\hat{Q}_1] $$
The Casimir is written in terms of these
second order operators as:
$$
C=4C_1+2C_2+4C_3+3\hat{Q}_1-3\hat{Q}_2-3
\hat{Q}_3 $$
and the Hamiltonian is:
$$ H= -Q_1+Q_2+Q_3 +{\rm constant}$$
where $Q_i$ are the conserved
quantities in spherical coordinates:
\begin{eqnarray*}
Q_1&=&{1\over 2}p_{\phi_1}^2+{m_0^2\over
\cos^2\phi_1}+ {m_1^2\over
\sin^2\phi_1}\\
Q_2&= &\tanh^2 \phi_2\left({1\over
2}p_{\phi_1}^2
\sin^2\phi_1 +{m_0^2\over
\cos^2 \phi_1}\right)+\cos^2 \phi_1
\left({1\over 2}p_{\phi_2}^2+{m_2^2\over
\tanh^2 \phi_2}\right)\\ 
&&- {1\over 2}p_{\phi_1}p_{\phi_2}\sin 2\phi_1
\tanh \phi_2\\
Q_3&= &\tanh^2 \phi_2\left({1\over
2}p_{\phi_1}^2
\cos^2\phi_1 +{m_1^2\over
\sin^2 \phi_1}\right)+\sin^2 \phi_1
\left({1\over 2}p_{\phi_2}^2+{m_2^2\over
\tanh^2 \phi_2}\right)\\
&&+ {1\over 2}p_{\phi_1}p_{\phi_2}\sin 2\phi_1
\tanh \phi_2
\end{eqnarray*}

The HJ equations corresponding to the
Hamiltonian (\ref{hamcom}) are: 
\begin{eqnarray} 
{1\over 2} \left( {\partial
S_1\over \partial
\phi_1}\right)^2+{m_0^2\over
\cos^2\phi_1} + {m_1^2\over
\sin^2\phi_1} & = & \alpha_1,\\ 
{1\over 2} \left( {\partial S_2\over \partial
\phi_2}\right)^2 +  {m_2^2\over
\sinh^2\phi_2}-  {\alpha_1\over
\cosh^2\phi_2} & = & \alpha_2.
\end{eqnarray}
The first one is the same as we got in $su(3)$
(\ref{hjsu3}), $\alpha_1$ is always positive
and $\alpha_2=E$.

The solutions depend on the values of the
parameters and energy

\noindent i) $E<0$
\begin{equation}
u_2={1\over 2|E|}
\left[-b_2+\sqrt{b_2^2+4\alpha_1E} \cos
2\sqrt{2|E|}t\right].
\end{equation}   
ii)
$E=0$ 
\begin{equation}
u_2={\alpha_1\over \alpha_1-m_2^2}+
2(\alpha_1-m_2^2) t^2.
\end{equation} 
iii) $E>0$
\begin{equation}
u_2={1\over 2E}\left[b_2+
\sqrt{b_2^2+4\alpha_1E}\cosh 2\sqrt{2E}t\right].
\end{equation}   
where $u_2=\cosh^2\phi_2$, $b_2=
E-\alpha_1+m_2^2$. The other equation can be
solved as we did in the previous cases. The
result is:
\begin{eqnarray} 
u_1 & = & {1\over
2\alpha_1}\left[ b_1+{1\over u_2}
\left[{b_1^2-4\alpha_1 m_0^2\over 
b_2^2+4\alpha_1E}\right]^{1/2}\left(
-(b_2u_2+2\alpha_1)\sin
2\sqrt{2\alpha_1}\beta_1\right.\right.
\nonumber\\ 
&&\left.\left. +2\sqrt{\alpha_1}
[( Eu_2-b_2)u_2-\alpha_1]^{1/2}
\cos2\sqrt{2\alpha_1}\beta_1\right)\right]
\end{eqnarray} 
where $u_1=\cos^2\phi_1$ and
$b_1=\alpha_1+m_0^2-m_1^2$.

\medskip

{\it II. Noncompact Cartan subalgebra (NC)}
\smallskip

There is only one noncompact Cartan
subalgebra. A representative can be chosen
according to the same criteria we used in
$su(1,1)$ (\ref{basesu11}), keeping one
element compact and the other (as in
$su(1,1)$) noncompact:
\begin{equation}
\left(\begin{array}{rrr} 
2i\\ & -i\\ &  & -i
\end{array}\right),\qquad
\left(\begin{array}{rrr} 
0 \\ & 0 & i\\ & -i& 0
\end{array}\right)
\end{equation}
and the basis for the corresponding MASA of
$u(2,1)$ will be:
\begin{equation}
Y_0=
\left(\begin{array}{rrr} 
i\\ & 0 \\ &  & 0 
\end{array}\right),\qquad
Y_1=
\left(\begin{array}{rrr} 0
\\  & i\\ &  & i
\end{array}\right),\qquad
Y_2=
\left(\begin{array}{rrr} 
0 \\ & 0 &
i\\ & -i& 0
\end{array}\right).
\end{equation}

The coordinates are as in $su(1,1)$:
\begin{eqnarray}
y_0 & = &  e^{ix_0}s_0,\nonumber\\ 
y_1 & = & e^{ix_1}(s_1\cosh x_2 +is_2\sinh
x_2),\\  
y_2 & = &  e^{ix_1}(-is_1\sinh x_2
+ s_2\cosh x_2).\nonumber
\end{eqnarray}

The Hamiltonian is:
\begin{eqnarray}
H&=&c\left({1\over 2}\left(
p_0^2+p_1^2-p_2^2\right)
+V(s)\right),\nonumber\\ 
V(s)&=&{m_0^2\over s_0^2}
+{(m_1^2-m_2^2)(s_1^2-s_2^2)
+4m_1m_2s_1s_2\over
(s_1^2+s_2^2)^2}\end{eqnarray}
where we will take $m_0>0$ and $m_1,m_2$ can
take any value, and the coordinates satisfy the
constraint (the same for all $su(2,1)$ MASAs, as
we have chosen the same metric in all cases):
$s_0^2+s_1^2-s_2^2=1$.

There are two systems of coordinates in
which the associated HJ equation separates, 
hyperbolic and complex elliptic \cite{BK2}.
We will use hyperbolic coordinates, defined
as:
\begin{eqnarray}
s_0 & = & \cosh \phi_2 \nonumber\\ 
s_1 & = & \sinh \phi_2
\sinh\phi_1\label{hyp}\\ 
 s_2 & = & \sinh
\phi_2 \cosh \phi_1\nonumber
\end{eqnarray} 
and the new Hamiltonian is ($c=-1$):
\begin{eqnarray}
H & = & {1\over 2}\left(
p_{\phi_2}^2-{p_{\phi_1}^2\over \sinh^2
\phi_2}\right)+V(\phi_1,\phi_2),\nonumber\\ 
V(\phi_1,\phi_2) & =&  -{m_0^2\over
\cosh^2\phi_2}+ {1\over \sinh^2\phi_2}
\left({m_1^2-m_2^2-2m_1 m_2 \sinh 2\phi_1
\over \cosh^2 2\phi_1}\right).
\end{eqnarray}
Note that the potential follows the same pattern
as the corresponding case in $su(1,1)$. It is
regular inside the domain:
$-\infty<\phi_1<\infty$, $0<\phi_2<\infty$,
and has also a saddle point at:
$\sinh 2\phi_1=-m_2/m_1$, $\tanh
\phi_2=\sqrt{|m_1/m_0|}$, when
$|m_1|<|m_0|$. As in the previous case, the
associated dynamical system has a center and
the trajectories in a neighborhood of it are
periodic.  

The basis for this MASA is $\{2X_1+X_2,X_8\}$,
and the corresponding second order elements
are:
$$C_1=(2X_1+X_2)^2,\qquad
C_2=\{2X_1+X_2,X_8\},\qquad C_3=X_8^2$$

The second order conserved quantities,
commuting with $2X_1+X_2$ and $X_8$, and
belonging to the enveloping algebra of
$su(2,1)$ are:
\begin{eqnarray}
\hat{Q}_1&=&X_2^2-X_7^2,\nonumber\\
\hat{Q}_2&=&X_3^2+X_4^2-X_5^2-X_6^2,\\
\hat{Q}_3&=&\{X_3,X_5\}+\{X_4,X_6\}\nonumber
\end{eqnarray}
with commuting relations:
$$ [\hat{Q}_3,\hat{Q}_1]=[\hat{Q}_2,\hat{Q}_3],
\qquad [\hat{Q}_1,\hat{Q}_2]=0 $$
The Casimir is written as:
$$ C=C_1-3C_3+3\hat{Q}_1+3\hat{Q}_2 $$
and the Hamiltonian is:
$$ H=Q_1+Q_2+{\rm constant}$$

Finally, the conserved quantities are expressed
in hyperbolic coordinates by:
\begin{eqnarray*}
Q_1&=&{1\over
2}p_{\phi_1}^2-{m_1^2-m_2^2-2m_1m_2\sinh 2
\phi_1\over\cosh^2 2\phi_1}\\
Q_2&=&{1\over 2} p_{\phi_2}^2-{m_0^2\over
\cosh^2\phi_2} - {1\over \tanh^2\phi_2}
\left( {1\over 2} p_{\phi_1}^2
-{m_1^2-m_2^2-2m_1m_2\sinh 2\phi_1\over 
\cosh^2 2\phi_1}\right)\\
Q_3& = &{1\over 2}\sinh 2\phi_1
\left(p_{\phi_2}^2+{1\over \tanh^2\phi_2}
p_{\phi_1}^2\right) -{\cosh 2\phi_1\over
\tanh\phi_2} p_{\phi_1}p_{\phi_2}\\
&&+m_0^2\tanh^2 \phi_2\sinh 2\phi_1 -
{(m_1^2-m_2^2)\sinh 2\phi_1 +2 m_1m_2\over
\tanh^2 \phi_2\cosh^2 2\phi_1}
\end{eqnarray*}

The Hamilton-Jacobi equation separates into two
ordinary differential equations:
\begin{eqnarray} {1\over 2}\left({\partial
S_1\over
\partial\phi_1}\right)^2-{m_1^2-m_2^2-2m_1m_2
\sinh 2\phi_1 \over\cosh^2 2\phi_1} & = &
\alpha_1, \\ {1\over 2}\left({\partial S_2\over
\partial\phi_2}\right)^2-{m_0^2\over
\cosh^2\phi_2}- {\alpha_1\over
\sinh^2\phi_2} & = & E.\label{nocomp} 
\end{eqnarray}
The solutions have the same form we have found
before.

\noindent i) $E<0$
\begin{equation}u_2={1\over 2|E|}\left[b_2+
\sqrt{b_2^2-4\alpha_1E}\cos
2\sqrt{2|E|}t\right].
\end{equation}   
ii) $E=0$ 
\begin{equation}u_2=-{\alpha_1\over
\alpha_1+m_0^2}+2(\alpha_1+m_0^2) t^2.
\end{equation} 
iii) $E>0$
\begin{equation}u_2={1\over 2E}\left[-b_2+
\sqrt{b_2^2-4\alpha_1E}\cosh 2\sqrt{2E}t\right]
\end{equation}  
where $u_2=\sinh^2\phi_2$,
$b_2=E+\alpha_1+m_0^2$.

The solution for the other coordinate is
obtained in the same way (with the change
$u_1=\sinh 2\phi_1$). The equation for this
coordinate is the same as that given in formula
(\ref{hjnc}) and its solutions can be found in
formulas (\ref{hjnc1}-\ref{hjnc5}). Due to the
possible different signs of the energy and the
constant $\alpha_1$, one should take care of
the square roots appearing in all the formulas.

\medskip

{\it III. Orthogonally decomposable
subalgebra (OD)}
\smallskip

The orthogonally decomposable subalgebra (a
representative of the class) has a basis
formed by a compact element and a nilpotent
one:
\begin{equation}
\left(\begin{array}{rrr} 
2i\\ & -i\\ &  & -i
\end{array}\right),\qquad
\left(\begin{array}{rrr} 
0 \\ & i& i \\ & -i&-i
\end{array}\right)
\end{equation}
and the basis for the corresponding MASA of
$u(2,1)$ is:
\begin{equation}
Y_0=\left(\begin{array}{rrr} 
i\\ & 0 \\ &  & 0 
\end{array}\right),
\qquad
Y_1=
\left(\begin{array}{rrr} 
0 \\  &i\\ &  & i
\end{array}\right),\qquad
Y_2=
\left(\begin{array}{rrr} 
0 \\ & i& i\\ & -i& -i
\end{array}\right).
\end{equation}

The coordinates have also a similar form to
those in $su(1,1)$ (\ref{nilsu11}):
\begin{eqnarray} 
y_0 & = & e^{ix_0}s_0,\nonumber\\ 
y_1 & = & 
e^{ix_1}((1+ix_2)s_1 +ix_2s_2),\\
 y_2 & = & 
e^{ix_1}(-ix_2s_1 + (1-ix_2)s_2).\nonumber
\end{eqnarray}

The Hamiltonian is:
\begin{eqnarray}
H &=&c\left({1\over 2}\left(
p_0^2+p_1^2-p_2^2\right)
+V(s)\right),\nonumber\\
V(s)&=&{m_0^2\over s_0^2}
-{m_2^2(s_1-s_2)\over (s_1+s_2)^3}
+{2m_1m_2\over (s_1+s_2)^2}
\end{eqnarray}
with $s_0^2+s_1^2-s_2^2=1$.

There are four coordinate systems associated
to this subalgebra, hyperbolic, horospheric,
elliptic parabolic and hyperbolic parabolic
\cite{BK2}. We will use again the hyperbolic
ones, defined as in (\ref{hyp}).

The Hamiltonian is ($c=-1$):
\begin{eqnarray}
H & = & {1\over 2}\left(
p_{\phi_2}^2-{p_{\phi_1}^2\over
\sinh^2
\phi_2}\right) +V(\phi_1,\phi_2),\nonumber\\
V(\phi_1,\phi_2) & = &
-{m_0^2\over\cosh^2\phi_2}- {1\over\sinh^2
\phi_2}\left( m_2^2e^{-4\phi_1}+
2m_1m_2e^{-2\phi_1}\right)
\end{eqnarray}

The potential is regular inside the domain:
$-\infty<\phi_1<\infty$,
$0<\phi_2<\infty$, and, as it happens in all
the $su(2,1)$ cases, has a saddle point at:
$\phi_1=(1/2)\log(|m_2/m_1|)$, $\tanh
\phi_2=\sqrt{|m_1/m_0|}$, when
$|m_0|>|m_1|, m_1m_2<0$. The situation is the
same as in all other cases in $su(2,1)$.

The second order operators in the enveloping
algebra of the MASA under consideration are
given by:
 $$C_1=(2X_1+X_2)^2,\qquad
C_2=\{2X_1+X_2,X_2+X_8\},\qquad
C_3=(X_2+X_8)^2 $$
and the quadratic constants of
motion:
\begin{eqnarray}
\hat{Q}_1&=&X_3^2+X_4^2-X_5^2-X_6^2,\nonumber\\
\hat{Q}_2&=&(X_3+X_5)^2+(X_4+X_6)^2,\\
\hat{Q}_3&=&X_7^2+2\{X_1,X_2+X_8\}\nonumber
\end{eqnarray}
satisfy the commutation relations:
$$[\hat{Q}_1,\hat{Q}_2]=[\hat{Q}_2,\hat{Q}_3],
\qquad [\hat{Q}_1,\hat{Q}_3]=0 $$
The Casimir is given in terms of these operators
by
$$C=C_1+3C_2-3C_3+3\hat{Q}_1-3\hat{Q}_3 $$ 
and the Hamiltonian is:
$$ H= -Q_1+Q_3 +{\rm constant}$$
We can write the conserved quantities in
hyperbolic coordinates:
\begin{eqnarray*}
Q_1&=&{1\over\tanh^2
\phi_2}\left({1\over
2}p_{\phi_1}^2+m_2^2e^{-4\phi_1}+2m_1m_2
e^{-2\phi_1}\right)-\left( {1\over
2}p_{\phi_2}^2-{m_0^2\over\cosh^2
\phi_2}\right)\\
Q_2&=&e^{2\phi_1}\left({1\over
2}p_{\phi_2}^2+m_0^2\tanh^2\phi_2 + {1\over
\tanh^2\phi_2}\left({1\over 2}
p_{\phi_1}^2+m_2^2
e^{-4\phi_1}\right)\right.\\
&&\left. -{1\over \tanh \phi_2}
p_{\phi_1}p_{\phi_2}\right)\\ 
Q_3&=&{1\over 2}p_{\phi_1}^2+m_2^2e^{-4\phi_1}+
2m_1m_2e^{-2\phi_2}
\end{eqnarray*}

The HJ equation is separated into the following
equations \begin{eqnarray}
{1\over 2}\left({\partial S_1\over
\partial\phi_1}\right)^2+\left(m_2^2
e^{-4\phi_1}+2m_1m_2e^{-2\phi_1}\right) & =&
\alpha_1,\label{jacodduno}\\
{1\over 2}\left({\partial S_2\over
\partial\phi_2}\right)^2-{m_0^2\over
\cosh^2\phi_2}- {\alpha_1\over
\sinh^2\phi_2} & = & E\label{jacoddos}.
\end{eqnarray}

Equation (\ref{jacoddos}) is integrated using
$u_2=\sinh^2\phi_2$. The result is the same
as in the previous case (\ref{nocomp}).
Equation (\ref{jacodduno}) is solved using
$u_1=e^{2\phi_1}$ (the same  change we use
in the nilpotent MASA of the $su(1,1)$
case), and the results are essentially the
same we have found above (see \ref{nileq}).

\medskip

{\it IV. Nilpotent subalgebra (NIL)}
\smallskip

The nilpotent subalgebra has a basis formed
by two nilpotent elements (one of order 2
and the other of order 3):
\begin{equation}
\left(\begin{array}{rrr} 
0 \\ & i& i\\ & -i& -i
\end{array}\right),\qquad
\left(\begin{array}{rrr} 
0  & i& i\\ i& 0 & 0\\
-i& 0 & 0
\end{array}\right)
\end{equation}
and the basis for the  MASA of $u(2,1)$ can be
obtained adding to these two matrices the
identity times the imaginary unit.

The new coordinates are defined through:
\begin{eqnarray}
y_0 & = & e^{ix_0}(s_0+i
x_2(s_1+s_2)),\nonumber\\ 
y_1 & = &  e^{ix_0}\left (i
x_2s_0+\left(1-{x_2^2
\over 2}+ix_1 \right)s_1 +\left(-{x_2^2\over
2} +ix_1\right)s_2\right),\\  
y_2 &= &e^{ix_0}\left(-ix_2s_0+
\left({x_2^2\over 2}
-ix_1\right)s_1+\left(1+{x_2^2 \over 2}-ix_1
\right)s_2 \right).\nonumber
\end{eqnarray}

The Hamiltonian is:
\begin{eqnarray}
H&=&c\left({1\over 2}\left(
p_0^2+p_1^2-p_2^2\right)
+V(s)\right),\nonumber\\ 
V(s)&=& {2m_0m_1+m_2^2\over (s_1^2+s_2)^2}-
{4m_1m_2s_0\over (s_1^2+s_2)^3}+
{m_1^2(4s_0^2-1) \over (s_1^2+s_2)^4}
\end{eqnarray}
with the constraint: $s_0^2+s_1^2-s_2^2=1$.

We have now two separable coordinate systems:
horospheric and semicircular parabolic
\cite{BK2}. We will use now the horospheric
ones, defined by:
\begin{equation}
s_0 = \phi_1 e^{\phi_2},\qquad
s_1 = \cosh \phi_2 -{1\over
2} \phi_1^2 e^{\phi_2},\qquad 
s_2 = \sinh \phi_2 +{1\over 2} \phi_1^2
e^{\phi_2}.
\end{equation}

The Hamiltonian is ($c=-1$):
\begin{eqnarray}
H &=&  {1\over 2}\left(
p_{\phi_2}^2-e^{-2\phi_2}
p_{\phi_1}^2\right) +
V(\phi_1,\phi_2),\nonumber\\
V(\phi_1,\phi_2) & = &  m_1^2
e^{-4\phi_2}-e^{-2\phi_2}
\left(m_2^2+2m_0m_1+4m_1\phi_1(
m_1\phi_1-m_2)\right).
\end{eqnarray}

The potential has no singularity in the whole
plain $(\phi_1,\phi_2)$. It has a saddle point
at: $\phi_1=m_2/2m_1$, $\phi_2=(1/2)\log
(m_1/m_0)$, when $m_0m_1>0$.
The situation is the same as in all other
cases in $su(2,1)$.

The second order elements in the enveloping
algebra of the nilpotent subalgebra are:
$$ C_1=(X_2+X_8)^2,\qquad
C_2=\{X_2+X_8,X_4+X_6\},\qquad
C_3=(X_4+X_6)^2 $$
and the constants of motion
\begin{eqnarray}
\hat{Q}_1 &=& 
3(X_3+X_5)^2-2\{2X_1+X_2,X_2+X_8\},\nonumber\\
\hat{Q}_2 &=& 
\{2X_1+X_2,X_4+X_6\}+6\{X_4,X_2+X_8\}-3
\{X_7,X_3+X_5\},\\
\hat{Q}_3 &=& 
4X_1^2+3X_2^2-2\{X_1,X_2\}+6X_3^2+6X_4^2-
3X_7^2 -\{4X_1-X_2,X_8\}\nonumber\\ 
&&+3\{X_3,X_5\}+3\{X_4,X_6\}\nonumber
\end{eqnarray}
have the following commutation relations:
$$
[\hat{Q}_1,\hat{Q}_2]=[\hat{Q}_3,
\hat{Q}_2],\qquad [\hat{Q}_1,\hat{Q}_3]=0 $$
The Casimir is
$$
C=-3C_1-3C_3-\hat{Q}_1+\hat{Q}_3 $$ 
and the Hamiltonian.:
$$ H= Q_1-Q_3 +{\rm constant}$$

Finally, the second order constant of motion
are given in horospheric coordinates by the
following expressions:
\begin{eqnarray*}
Q_1&=&{1\over 2}
p_{\phi_1}^2+4m_1\phi_1(m_1\phi_1-m_2)\\
Q_2&=&{1\over 2}\phi_1p_{\phi_1}^2-{1\over
2}p_{\phi_1}p_{\phi_2}+(m_2^2+2m_0m_1)\phi_1\\
&&- m_1e^{-2\phi_2}(2m_1\phi_1-m_2)
+4m_1\phi_1^2(m_1\phi_1-m_2)\\
Q_3&=&(1+e^{-2\phi_2})\left({1\over
2}p_{\phi_1}^2+4m_1\phi_1(m_1\phi_1-m_2)
\right) \\
&&- \left({1\over 2}
p_{\phi_2}^2+m_1^2e^{-4\phi_2}-(m_2^2+
2m_0m_1)e^{-2\phi_2}\right)
\end{eqnarray*}

This is the most interesting case, in the
sense that the others are easily reduced to
the cases in dimension 1, while this
nilpotent subalgebra does not appear in the
$su(1,1)$ case. However, the solutions are
still very similar to those found before. It
is worth mentioning here, that all the
potentials we have construct (and any
potential we could construct by using this
method) are always inverse quadratic
potentials in the coordinates, and the
solutions have always similar forms (though
they depend on the specific characteristics
of these potentials and the constants
involved).

The HJ equation is separated into the
following equations
\begin{eqnarray}
{1\over 2}\left({\partial S_1\over
\partial\phi_1}\right)^2+m_2^2
+2m_0m_1+4m_1\phi_1(m_1\phi_1-m_2) & =& 
\alpha_1,\\
{1\over 2}\left({\partial S_2\over
\partial\phi_2}\right)^2+ m_1^2
e^{-4\phi_2}-\alpha_1e^{-2\phi_2} & = & E. 
\end{eqnarray}

The change $u_2=e^{2\phi_2}$ allows to solve
the second equation, and the other one is
solved directly. The solutions are:

\noindent i) $E<0$
\begin{equation}
u_2={1\over 2|E|}\left[\alpha_1+
\sqrt{\alpha_1^2+4m_1^2E}\cos
2\sqrt{2|E|}t\right].
\end{equation}   
ii) $E=0$ 
\begin{equation}
u_2={m_1^2\over \alpha_1}+2\alpha_1 t^2.
\end{equation}
iii) $E>0$
\begin{equation}
u_2={1\over 2E}\left[-\alpha_1+
\sqrt{\alpha_1^2+4m_1^2E}\cosh
2\sqrt{2E}t\right].
\end{equation}  

The first equation gives the value of the
$\phi_1$ coordinate:
\begin{eqnarray}
\phi_1& = & {1\over 2m_1}\left[ m_2+{1\over
u_2} \left[{\alpha_1-2m_0m_1\over 
\alpha_1^2+4m_1^2E}\right]^{1/2}\left(
(\alpha_1u_2-2m_1^2)\sin
2\sqrt{2}\beta_1m_1\right.\right.\nonumber\\
&&\left.\left. +2m_1[(Eu_2
-\alpha_1)u_2-m_1^2]^{1/2}
\cos2\sqrt{2}\beta_1m_1\right)\right].
\end{eqnarray}

In Table II, we present a summary of
these Hamiltonians in the two dimensional
case.

\sect{Conclusions}

We have presented in this work a complete
analysis of a series of one and two dimensional
integrable Hamiltonians, which in the 2-dimensional
case are superintegrable in the sense described in the
Introduction. Though the one-dimensional case is
always an integrable system, let us remark the
importance and applications of the potentials described
in Section 2. Regarding the two dimensional ones, we
have provided them with a set of conserved quantities
which allows to study the HJ equations in several
separable coordinate systems and compute in some
interesting cases the explicit solutions. 

The one-dimensional Hamiltonians obtained
here have been extensively studied in the
literature from many other points of view. See
for instance \cite{MQ} for a recent application
of Morse potentials. As an example of these
different approaches, all of them appear in the
classification of quasi-exactly solvable
Schr\"odinger operators \cite{Tu,GK}, as
particular types of these systems corresponding
to the so called exactly solvable systems. 
Following the classification in \cite{GK}, the
exactly solvable potentials of Cases 1 and 2 are
just the ones we have obtained associated
to  $su(1,1)$ and its compact and noncompact
Cartan subalgebras. The first one (\ref{pot1})
is the celebrated P\"oschl-Teller potential.
That appearing in case 3 is the Morse potential
(\ref{pot2}), which we get using the nilpotent
subalgebra of $su(1,1)$. Finally the potential
(\ref{pot3}), associated to the Cartan
subalgebra of $su(2)$ is related to the
modified harmonic oscillators appearing in
\cite{GK} as cases 4 and 5. One has to take
into account in this case, that QES
potentials, as studied in \cite{GK}, are
defined in the line (or half-line), and we
are working here in a sector of $S^1$ (see
also \cite{LM} for a study of harmonic
oscillators in a sector). The relation is
not surprising at all if one considers that
QES systems in the line are related to the
complex Lie algebra $sl(2)$ \cite{Tu}, and
the ones we get here reflect the invariance
under $su(2)$ and $su(1,1)$, the real forms
of $sl(2)$. In the QES setting,
Schr\"odinger operators belong to the
enveloping algebra of a Lie algebra, while
in our approach, the corresponding classical
Hamiltonians are second order Casimirs of
the algebra, and hence, they are particular
cases (exactly solvable) of the former.

Two prolongations of this study are now in
progress. One of them is the use of
contractions in Lie algebras to obtain other
Hamiltonian systems associated to different
algebras, not necessarily semisimple. In
this sense, the Hamiltonians, the conserved
quantities and the coordinate systems can be
obtained by contraction \cite{IP,CO2}. The
second one is to apply this approach to the
quantum case, considering the Schr\"odinger
equation with these potentials \cite{CN}. We also plan
to study the links of this theory with QES
systems and the possibility of considering
partial integrability and partial variable
separation in HJ equations \cite{Ha}.

\section*{Acknowledgments}
 
The authors would like to thank J. Negro for
many interesting discussions. This work has
been partially supported by CICYT (Spain,
Projects  PB95-0719 and PB95-0401) and Junta
de Castilla y Le¢n (Spain).

\vfill\eject

\begin{center}
TABLE I. One-dimensional potentials
\vskip 0.5cm
\begin{tabular}{|l|l|c|l|}
\hline
Algebra & MASA &
Kinetic term & Potential\\
\hline 
&&&\\
$su(2)$ & Compact Cartan & $p_{\phi}^2$ & 
$\displaystyle{{m_0^2\over
\cos^2\phi}+{m_1^2\over
\sin^2\phi}}$\\ &&&\\
\hline &&&\\
$su(1,1)$ &  Compact Cartan & $p_{\phi}^2$ & 
$\displaystyle{-{m_0^2\over
\cosh^2 \phi}+{m_1^2\over
\sinh^2 \phi}}$\\ &&&\\ & Noncompact Cartan
& $p_{\phi}^2$ & 
$\displaystyle{-{m_0^2-m_1^2 + 2m_0m_1 \sinh
2\phi \over \cosh^2 2\phi}}$\\ &&&\\ & 
Nilpotent & $p_{\phi}^2$ & 
$\displaystyle{m_1^2 e^{-4\phi} - 2m_0m_1
e^{-2\phi}}$\\ &&&\\
\hline
\end{tabular}
\end{center}

\begin{center}
TABLE II. Two-dimensional potentials
\vskip 0.5cm
\begin{tabular}{|l|c|l|l|}
\hline
Algebra & MASA &
Kinetic term & Potential\\
\hline &&&\\
$su(3)$ & CC & ${1\over 2} \left( 
{\scriptstyle
p_{\phi_2}^2}+{p_{\phi_1}^2\over
\cos^2 \phi_2}\right)$ & 
${ {1\over
\cos^2\phi_2}\left({m_0^2\over
\cos^2\phi_1}+{m_1^2\over
\sin^2\phi_1}\right)+ {m_2^2\over
\sin^2\phi_2}}$\\ &&&\\
\hline &&&\\
$su(2,1)$ &  CC & ${1\over 2}
\left({\scriptstyle
p_{\phi_2}^2}-{p_{\phi_1}^2\over
\cosh^2 \phi_2}\right)$ & 
${ -{1\over
\cosh^2\phi_2}\left({m_0^2\over
\cos^2\phi_1}+  {m_1^2\over
\sin^2\phi_1}\right)+ {m_2^2\over
\sinh^2\phi_2}}$\\ &&&\\ & NC & ${1\over
2}\left( {\scriptstyle
p_{\phi_2}^2}-{p_{\phi_1}^2\over
\sinh^2
\phi_2}\right)$ & 
${ -{m_0^2\over \cosh^2\phi_2}+ {1\over
\sinh^2\phi_2}\left({m_1^2-m_2^2-2m_1 m_2
\sinh 2\phi_1
\over \cosh^2 2\phi_1}\right)}$\\ &&&\\ & 
OD & ${1\over 2}\left( {\scriptstyle
p_{\phi_2}^2}-{p_{\phi_1}^2\over
\sinh^2 \phi_2}\right)$ & 
$ -{m_0^2\over\cosh^2\phi_2}-
{1\over\sinh^2\phi_2}{\scriptstyle \left(
m_2^2e^{-4\phi_1}+
2m_1m_2e^{-2\phi_1}\right)}$\\  &&&\\  & 
NIL& ${1\over 2}\left({\scriptstyle  
p_{\phi_2}^2-e^{-2\phi_2}
p_{\phi_1}^2}\right)$ & 
${\scriptstyle  m_1^2
e^{-4\phi_2}-e^{-2\phi_2}
\left(m_2^2+2m_0m_1+4m_1\phi_1(
m_1\phi_1-m_2)\right) }$\\ &&&\\
\hline
\end{tabular}
\end{center}

\end{document}